\newcommand{\beq}{\begin{equation}}
\newcommand{\eeq}{\end{equation}}
\newcommand{\bea}{\begin{eqnarray}}
\newcommand{\eea}{\end{eqnarray}}

\newcommand{\gsim}{\lower.7ex\hbox{$\;\stackrel{\textstyle>}{\sim}\;$}}
\newcommand{\lsim}{\lower.7ex\hbox{$\;\stackrel{\textstyle<}{\sim}\;$}}

\documentclass[aps,prev,preprintnumbers,floatfix,nofootinbib,11pt]{revtex4-1}
\usepackage{graphicx}
\usepackage{epstopdf}
\usepackage{mathrsfs}
\usepackage{amssymb}
\usepackage{verbatim}
\usepackage{color}
\usepackage{multirow}
\usepackage{amsmath}
\usepackage{float}
\usepackage[normalem]{ulem}
\usepackage[lofdepth,lotdepth,caption=false]{subfig}

\def\stacksymbols #1#2#3#4{\def\theguybelow{#2}
    \def\vp{\lower#3pt}
    \def\sp{\baselineskip0pt\lineskip#4pt}
    \mathrel{\mathpalette\intermediary#1}}

\def\intermediary#1#2{\vp\vbox{\sp
     \everycr={}\tabskip0pt
     \halign{$\mathsurround0pt#1\hfil##\hfil$\crcr#2\crcr
              \theguybelow\crcr}}}

\def\be{\begin{equation}}
\def\ee{\end{equation}}
\def\bea{\begin{eqnarray}}
\def\eea{\end{eqnarray}}
\begin{document}

\vspace*{1mm}

\title{Discovery potential of FASER$\nu$ with contained vertex and through-going events }

\author{Pouya Bakhti$^{a}$}
\email{pouya$_$bakhti@ipm.ir}
\author{Yasaman Farzan$^{a}$}
\email{yasaman@theory.ipm.ac.ir}
\author{Silvia Pascoli$^{b}$}
\email{silvia.pascoli@durham.ac.uk}
\vspace{0.2cm}

\affiliation{
${}^a$School of physics, Institute for Research in Fundamental Sciences (IPM),\\
P.O. Box 19395-5531, Tehran, Iran
\\
${}^b$ 
Institute for Particle Physics Phenomenology, Department of Physics, Durham University, Durham
DH1 3LE, U.K.}

\begin{abstract}  FASER$\nu$  is a newly proposed detector whose main mission is to detect the neutrino flux from the collision of the proton beams at the ATLAS Interaction Point (IP) during  the run III of the LHC in 2022-2024. We show that this detector can also test certain beyond standard model scenarios, especially the ones in which the neutrino interaction with matter fields can produce new unstable particles decaying back into charged leptons. Models of this kind are  motivated by the MiniBooNE anomaly.  We show that, if the new physics involves multi-muon  production by neutrinos scattering off matter fields, including  the neutrino flux interactions in the rock before the detector in the analysis ({\it i.e.,} accounting for the through-going muon pairs) can significantly increase the effective detector mass and its sensitivity to new physics. We  propose a concrete model that can give rise to such a multi-muon signal. 
\end{abstract}

\maketitle

\section{Introduction}

Neutrinos are copiously produced
at the interaction points of the LHC experiments. However, their detection at the main detectors of the LHC ({\it i.e.,} CMS, ATLAS, ALICE and LHCb) is not possible because of the large background from other particles produced at the Interaction Point (IP). To detect high energy neutrino flux from the IP, the FASER$\nu$ experiment \cite{Abreu:2019yak} has been proposed. It will take data during run III of the LHC at a distance of 480~m upstream of the ATLAS interaction point. The FASER$\nu$ detector is composed of layers of Tungsten interleaved with emulsion film with a total size of 25 cm$\times$25 cm$\times$1.3 m and a total mass of 1.2 ton.
As the neutrinos travel in a straight line and traverse 10 meters of concrete and 90 meters of soil, they can reach the detector.
FASER$\nu$ can detect the charged leptons produced in the Charged Current (CC) interactions of all three flavors of neutrinos inside the detector. In principle, the muons produced by the $\nu_\mu$ CC interactions in the rock and concrete before FASER$\nu$ can also give rise to a signal.
However, the background of muons from IP 
makes using these muons challenging.

As shown in \cite{Abreu:2019yak,YASI,POO,Felix}, FASER$\nu$ can also search for new physics. We focus in particular on the muon channel. For the first time, in this paper, we show that in the presence of multi-muon events originating from new physics, the effective mass of the detector can be increased by a factor of 4 by including events starting in the rock or concrete before the detector. This effect is similar to the enhancement of the effective volume of ICECUBE to
detect $\nu_\mu$ by searching for the through-going muons from below.
We point out that, if FASER$\nu$ is equipped with a scintillator detector that can record the arrival time of the muons comprising a multi-muon event,
the background from the pile-up of the muons penetrating from IP as well as from the charged current interaction of $\nu_\mu$ in the rock can be made negligible.

In this paper, we first show that with this technique it is possible to considerably enlarge the effective mass of the detector by registering the multi-muon events produced outside the detector.
We then introduce a model that leads to a signal of charged lepton pairs.
The model, which is also motivated by the MiniBooNE anomaly \cite{MiniBooNE}, includes a new right-handed neutrino, $N$, with a mass in the range of few 100 MeV to $\sim 10$ GeV and a dark photon, $Z'$ with a coupling converting $\nu$ and $N$ to each other \cite{Ballett:2018ynz, Bertuzzo:2018itn}. The well-known technique to obtain such off-diagonal $Z'$ interaction is to mix the new fermions with the Standard Model (SM) neutrinos. Since there are strong upper bounds on the mixing, obtaining large couplings for the off-diagonal $Z'$ interaction is usually a challenge but in the appendix, we introduce a new trick to circumvent the bounds and obtain a coupling as large as $0.01$.

With this coupling,
The scattering of the neutrino flux off nuclei will create $N$. If $N$ is heavier than $Z'$, it can decay into $Z'$ which in turn decays into $e^-e^+$ and, for heavy enough $Z'$, also into $\mu^-\mu^+$. If $N$ is lighter than $Z'$, it will go through three-body decays, producing $e^-e^+$ or $\mu^+\mu^-$ pairs. By studying the energy-momentum and tracks of the final charged leptons originated from a vertex inside the FASER$\nu$ detector, further information about the lifetime and masses of the new particles can be derived.
We also show how we can build a variation of the model in which the $N$ decay leads to a four muon signal instead of a dilepton one. We demonstrate how including the through-going four-muon signals can extend the capability of FASER$\nu$ to probe smaller values of the neutrino coupling to the new particles.

FASER$\nu$ will be located right before the FASER detector \cite{FASERhollow,faserHOLLOW} such that the neutrino beam enters FASER immediately after exiting the FASER$\nu$ detector. The natural question is whether by including FASER, the data sample can be increased. Unlike FASER$\nu$ which has a dense medium to detect neutrinos, FASER is hollow so  inside FASER, the number of events from neutrino scattering will be negligible. However, the muons produced inside the rock can reach FASER after traversing FASER$\nu$. Fortunately, FASER is equipped to record timing, too, so the possibility of enlarging the effective  volume of the detector by studying the through-going multi-muon events applies for FASER, too.

This paper is organized as follows.
In sect. \ref{Enhance}, we show under what conditions including the neutrino interactions in the rock and concrete can enlarge the effective volume
of the detector. In sect. \ref{model}, we outline the basics of the model that can lead to neutrino upscattering to $N$ and its subsequent decay that can yield a multilepton signal. The details of the model, with emphasis on the trick to obtain large upscattering cross-section for neutrinos and on obtaining the $\mu \bar{\mu}\mu \bar{\mu}$ signal, are presented in the appendix. In sect. \ref{decay}, we discuss the decay modes of $N$ and their signature at FASER$\nu$. In sect. \ref{production}, we show how to calculate the production rate of $N$ from the $\nu$ interaction and how to compute the number of multi-lepton events that can be registered at FASER$\nu$. The results on the number of events and the bounds that can be obtained by FASER$\nu$ and its upgrades are presented in sect. \ref{res}.
A summary and a discussion on the extension of the results to other theoretical models are given in sect. \ref{summary} with an emphasis on the role of time recording scintillator.

\section{Through-going and contained vertex multi-muon events and their backgrounds \label{Enhance}}
In this section, after reviewing backgrounds for multimuon events and suggesting strategies for reducing them, we discuss under what conditions the through-going multi-muon events can be invoked to discover new physics.
The only particle other than neutrinos that can travel in the rock up to the detector is the muon. Muons will lose energy in the rock so reconstructing their energy-momentum at the detector will not give much information. However, their directions remain fairly faithful to their initial direction at the production. In fact, at these energies and for these distances the deviation from the original direction will be less than $0.5^\circ$
(see the last paragraph of page 5 of Ref.
\cite{Antonioli:1997qw}).
The angular resolution of the FASER$\nu$ detector is much better than this value
\cite{Abreu:2019yak} and at the level of sub-milliradians so we can practically neglect the uncertainty due to the angular resolution.
Consider a muon pair produced at a distance of $d=14$ m from the detector. The two lines connecting the production point to the center and the edge of the detector of size $25~{\rm cm} \times 25~{\rm cm}$ subtend an angle of $[((0.25/2)/14) \times (180/\pi)]^\circ=0.5^\circ$.
Thus, roughly speaking, for the muon pair produced in the vicinity of 14 m from the detector (despite the deviation of the direction of the muon during the propagation) we can distinguish that they have originated in a distance between (7~m to 14~m) from the detector along the beam and are not coming directly from the Interaction Point.

According to
\cite{Abreu:2019yak}, during the run III of the LHC (2022-2024), $O(10^9)$ muons will reach FASER$\nu$ from the IP or the interaction of the neutrino flux in the rock before the detector. The probability that two separate events out of these $10^9$ muons pile up in a single bunch crossing ($\sim 1$ ns) is negligible. Thus, recording the timing of the multi-muon event will significantly reduce the background.
FASER$\nu$ is an emulsion detector that cannot record time but if, as it is suggested in \cite{Abreu:2019yak}, the detector is equipped with a scintillator plate, recording the timing of the arrival of the muons with nanosecond accuracy will become a reality. Still the dimuon events can originate from pair production from photons \cite{cern} or from charm production via $\nu_\mu$ CC interaction ({\it i.e.,} $\nu_\mu +{\rm nucleon}\to \mu +c+X$) and the subsequent leptonic decay of the charm. The number of such charm-induced dimuon events originating outside the detector can be estimated as 
$$B \sim \frac{\sigma (\nu_\mu +{\rm nucleus}\to \mu+c+X)}{\sigma (\nu_\mu +{\rm nucleus}\to \mu+X)}\times {\rm Br}(c\to \mu+...)\times\frac{{\rm soil \ and \ concrete \ mass   } }{{\rm FASER}\nu \ {\rm mass}}\times({\rm No\ of\ muons\ at\ FASER}\nu),$$
where ${\rm Br}(c\to \mu+...)\simeq 0.1$, the mass of soil and concrete can be estimated as $\rho_{soil}\simeq \rho_{con}\simeq 2.5$ ton/$m^3$ times the volume of a cone with a height of 100 meters and a base equal to one side of the FASER$\nu$ detector; {\it i.e.,}
$\rho_{soil}(100\times 0.25\times 0.25~{\rm m}^3)/3$. The number of muons from the charged current interaction of neutrinos  during the run III of the LHC will be about  20000 \cite{Abreu:2019yak}. From Fig. 19 of \cite{Abreu:2019yak}, we observe that $0.1<{\sigma (\nu_\mu +{\rm nucleus}\to \mu+c+X)}/{\sigma (\nu_\mu +{\rm nucleus}\to \mu+X)}<0.2$. Putting these pieces of information together, we find 
that  during the run III of the LHC, $800 <B<1.6~{\rm k}$.  Notice that the dimuon events from the photon pair production have to be added to this background. Another source of the background can be tri-muon events in the standard model \cite{Barger:1979eg} when one of the muons goes missing. As we shall discuss in sect. \ref{decay}, the background events with a vertex inside the detector can be suppressed by the measurement of the energy-momentum of the dimuon or by considering the event topology but these methods cannot be used for suppressing the background for dimuon events originating in the rock.
FASER$\nu$ can declare discovery of new physics with dimuon signal from the rock only if the number of events is much larger than the root of the number of the background events; {\it i.e.,} the number
of dimuons from new physics is much larger than $\sqrt{B}>30$.

If the new physics signal instead of dimuon is  four-muons, we do not need to worry about the background. If FASER$\nu$ records two  muon pair events within 1~ns, we can make sure that they originate from new physics even
if we do not reconstruct the directions of the muons. For such signal events,
the effective mass of the detector will be enhanced by including the events originating in the rock before the detector.

In the next section and in the appendix, we shall develop a model in which one or two pairs of forward going muons are produced by the interaction of the neutrino flux in the rock and concrete before the detector. We then demonstrate how invoking the through-going muon pairs will increase the sensitivity of FASER$\nu$ to search for new physics.

\section{The model \label{model}}
Refs.
\cite{Ballett:2018ynz, Bertuzzo:2018itn, Abdullahi:2020nyr} propose a novel idea to explain the low energy $e^-$ excess observed in the MiniBooNE experiment. It is based on introducing a sterile neutrino of mass of a few 100 MeV mixed with $\nu_\mu$ and a new light $U(1)$ gauge boson, $Z'$, coupled both to the quarks and to the neutrinos with a coupling of form
\be \label{off-coupling} g_{\nu N
}Z'_\alpha \bar{\nu}_\mu \gamma^\alpha \frac{1-\gamma_5}{2} N.\ee
Through this coupling, $\nu_\mu$ (or $\bar{\nu}_\mu$) can scatter off nuclei converting to $N$ (or to $\bar{N}$). In the models introduced in Refs.
\cite{Ballett:2018ynz, Bertuzzo:2018itn}, the coupling of $Z'$ to the quarks emerges because of the kinetic mixing between $Z'$ and the photon so $Z'$ also couples to $e^-e^+$. As a result, the produced $N$ subsequently decays producing an electron-positron pair via $N\to \nu (Z')^*\to \nu e^-e^+$~\footnote{ In the self-consistent implementation of this idea in Ref.~\cite{Abdullahi:2020nyr}, $N$ decays into a dilepton pair and a lighter heavy neutral lepton, which could further decay inside the detector into another dilepton pair and neutrino.} As shown in \cite{Ballett:2018ynz}, in the MiniBooNE experiment, these pairs can be misidentified as charged current interactions of $\nu_e$ or $\bar{\nu}_e$. In the case that $m_{Z'}>m_N$, as is assumed in \cite{Ballett:2018ynz}, the $N$ decay will be a three-body decay mediated by an off-shell $Z'$. On the other hand, in the case $m_{Z'}<m_N$, as is assumed in \cite{Bertuzzo:2018itn}, $N$ can decay into $Z' \nu$ and $Z'$ subsequently decays into the $e^-e^+$ pair.

In the model presented in \cite{Ballett:2018ynz}, $N$ is charged under the new $U(1)$ gauge symmetry so we would have a coupling of $g' \bar{N} \gamma^\alpha N Z'_{\alpha}$ where $g'$ is the gauge coupling which can be of order of 1. A mixing between $N$ and $\nu_\mu$ leads to the coupling of form (\ref{off-coupling}) with $g_{\nu N}=g' U_{\mu 4}$.
To be precise $\nu_\mu$ in Eq.~(\ref{off-coupling}) is not a flavor (or electroweak) eigenstate but a linear combination of light eigenstates as $\sum_{i=1}^3 |i\rangle \langle i |\nu_\mu \rangle$ that is produced in the processes with energy scale below the mass of $\nu_4$. This combination corresponds to the coherent state that is produced in the muon and pion decay. Since in this model $g_{\nu N}$ is proportional to the mixing ($g_{\nu N}=g'U_{\mu 4}$), for a given $m_N$, the bound on the mixing will be translated into a bound on $g_{\nu N}$.

Let us enumerate the bounds on the mixing of a sterile neutrino, $\nu_4$ of a mass of $m_4$ with active neutrinos.
For $m_4 <m_\pi -m_\mu \simeq 30$ MeV, the search for $\pi \to \mu \nu_4$ in the PIENU experiment \cite{Aguilar-Arevalo:2019owf} sets an upper bound of $10^{-3}$ on
$U_{\mu 4}$. On the other hand, for $70 ~{\rm MeV}<m_4 <m_K-m_\mu$ the searches for $K \to \mu \nu_4$ by KEK \cite{Hayano:1982wu}, by E949 \cite{Artamonov:2014urb} and by NA62 \cite{Estrada-Tristan:2019jyr} yield a very stringent
bound on $g_{\nu N}$.
For heavier $\nu_4$, the bound comes from NuTeV \cite{Vaitaitis:1999wq}. All the bounds are summarized in Ref. \cite{Dev}. However, for $ 30~ {\rm MeV}< m_4< 70~ {\rm MeV}$, the strongest bound on $U_{\mu 4}$, which comes from studying the muon decay spectrum \cite{MUspectrum}, is relatively relaxed and is of order of few$\times 10^{-2}$ \cite{Dev}.

As we demonstrate in the appendix, it is possible to build models for the coupling of form in Eq (\ref{off-coupling}) with nonzero $g_{\nu N}$ even with a vanishing mixing between $N$ and $\nu_\mu$. Throughout the paper, we, therefore, take $g_{\nu N}$ to be independent of $U_{\mu 4}$ and study its signatures at FASER$\nu$.
In addition to (or instead of) $\nu_\mu$, $\nu_e$ and/or $\nu_\tau$ could have a coupling of form (\ref{off-coupling}) to $Z'$ and $N$. We however focus on the case that the coupling is exclusively to $\nu_\mu$ because of two reasons: (i) The flux of $\nu_\mu$ at FASER$\nu$ is higher than those of $\nu_e$ and $\nu_\tau$; (ii) To keep connection with the solution to the MiniBooNE excess. However, similar analysis and discussion can be repeated for the $\nu_\tau$ or $\nu_e$ coupling to $N$ and $Z'$.

The basis of the toy model described in the appendix is introducing two sterile neutrinos $N'$ and $N$ with an off-diagonal coupling of the form $g' \bar{N} \gamma^\mu
N' Z'_\mu$. The mixing of $N$ with active neutrinos is taken to be zero or negligible. $N'$ has a mass in the range $(30~{\rm MeV},70~{\rm MeV} )$ so it can have a relatively large mixing with $\nu_\mu$ leading to a relatively large $g_{\nu N}$. We take $N$ to be heavier than $N'$ so that $N$ can decay into $N'$.
If the coupling of $Z'$ to the SM fermions is through its mixing to the photon, $\epsilon$, it will not couple to $\nu \bar{\nu}$ and as a result, at the tree level, the decay modes $N \to Z'^{(*)} N' \to N' \nu \bar{\nu}$ are closed but we can have
$$N \to N' Z'^{(*)} \to N' e^-e^+ \ \ {\rm or} \ \ N \to N' Z'^{(*)} \to N' \mu^-\mu^+ \ \ {\rm for} \ m_N-m_{N'}>2 m_\mu .$$
As discussed in the appendix, $Z'$ may also couple to a pair of new scalars, $a$ and $\bar{a}$, whose decays produce a pair of $\mu \bar{\mu}$. In this case, the $N$ decay can produce the background-free signal of two pairs of $\mu \bar{\mu}$
\be N \to N' Z'^{(*)} \to N' a \bar{a} \ \ {\rm followed ~ by} \ \ a\to \mu \bar{\mu} \ \ {\rm and} \ \ \bar{a}\to \mu \bar{\mu}.\label{mmmm}\ee
The $N'$ particle can decay via the electroweak interactions as $N' \to \nu_\mu Z^* \to \nu_\mu f \bar{f}$ where $f=e$ or $f =\nu$ with a decay rate of $\Gamma \sim G_F^2 m_{N'}^5 |U_{\mu 4}|^2/100 \pi^3$. Thus, an $N'$ with an energy of $E_{N'}$ will travel a distance of $\Gamma^{-1} (E_{N'}/m_{N'}) \sim 3 \times 10^7 ~{\rm m} |U_{\mu 4}|^{-2} (E_{N'}/100~{\rm GeV}) (50~{\rm MeV}/m_{N'})^6$ which is much larger than the distance between FASER$\nu$ and the interaction point.

We now have a model for the coupling of form in Eq. (\ref{off-coupling}).
In order to produce $N$ via scattering of $\nu$ off the nuclei, we also require $Z'$ to couple to the quarks.
Let us take the coupling of $Z'$ to the standard model fermions, $f$ as
$$ q_f' \bar{f} \gamma^\mu f Z'_\mu.$$
As discussed in \cite{Ballett:2018ynz}, this can be achieved by the following kinetic mixing between the photon field strength $F_{\mu \nu}$ and the field strength of the new $U(1)$ gauge symmetry, $Z'_{\mu\nu}$
$$- \frac{\epsilon}{2} Z'_{\mu \nu}F^{\mu \nu}.$$
Going to the canonical basis in which the kinetic terms are diagonal and properly normalized, all the charged fermions of the SM, $f$, will obtain a coupling to $Z'$ proportional to their electric charge: \be \label{relation}q_f'=e\epsilon q_f.\ee
For $20~{\rm MeV}<m_{Z'}<10$~GeV, the strongest bound on $\epsilon$ comes from BABAR with $\epsilon <7 \times 10^{-4}$ \cite{Lees:2014xha} (see also \cite{Archilli:2011zc,Merkel:2014avp}). For such values of $\epsilon$, the $Z'$ decay in the early universe will take place long before the neutrino decoupling so that there would be no deviation from the standard big bang nucleosynthesis prediction. If $Z'$ is lighter than a few hundred MeV, it can be produced via $\gamma +e^- \to e^-+Z'$ inside the supernova core and become thermalized until the temperature drops below the $Z'$ mass. Since the $Z'$ mean free path, as well as its decay length, is much shorter than the size of the core, the $Z'$ production will not dramatically affect the supernova core evolution.
If $N$ is heavier than a few 100 MeV, it cannot be produced inside the supernova core but it can change the mean free path of $\nu_\mu$ and $\bar{\nu}_\mu$ in the inner core via $\nu_\mu+\bar{\nu}_\mu \to Z' +Z'$. However, at the neutrinosphere with a temperature lower than the $Z'$ mass, this process cannot take place so the neutrino emission duration will not be significantly prolonged. $N'$ can be produced via $\nu_\mu Z' \to N' Z'$ with a rate of $\sim (g' g_{\nu N})^2 T^5/(4\pi m_N^4)$ and via the weak interactions with a rate of $\sim G_F^2 T^5 |U_{\mu 4}|^2/(4\pi)$. The produced $N'$ will be thermalized by scattering $Z'N' \to N' Z'$ via a $t$-channel $N$ exchange with a mean free path in the inner core smaller than that of the active neutrinos so it cannot transfer the energy of the core outside or affect the convection in the core. Once the core is depleted of neutrinos and the temperature drops below the $Z'$ mass, $N'$ cannot be reproduced and it decays within $\sim 0.01$ sec emitting $\nu_\mu$. With the current precisions, the supernova constraints cannot rule out the scenario but it can alter the $\nu_\mu$ spectrum which in the future can be tested.
Unless stated otherwise, throughout this paper, we shall assume that the coupling of the $Z'$ to the quarks and charged leptons is through kinetic mixing between $Z'$ and the photon so the relation in Eq. (\ref{relation}) holds.

Another option that we shall entertain in this paper is gauging the anomaly free combination
$$B-3L_\tau.$$
Then, $q'_e=q'_\mu=0$ and $q'_u=q'_d=q'_s=q'_c=q'_b=q'_\tau/9$.
We are interested in $Z'$ heavier than $4m_\mu$ coupled to $a \bar{a}$ such that its dominant decay mode is $Z' \to a\bar{a}$, leading to the $\mu \bar{\mu}\mu \bar{\mu}$ as in Eq. (\ref{mmmm}). Since in this option $Z'$ does not couple to $e$ and $\mu$, the bounds from KLOE \cite{KLOE} and BABAR \cite{Lees:2014xha, BAB} will be relaxed. As $Z'$ is heavier than 400 MeV and decays fast, bounds from the early universe or supernova cooling are irrelevant. Since $Z'$ couples  both to $\nu_\tau$ and to quarks, it can contribute to non-standard interaction of the tau neutrinos, $\epsilon_{\tau \tau}^{u(d)}=(q'_\tau q'_{u(d)}/m_{Z'}^2 )/(\sqrt{2} G_F)$ \cite{Felix}. Considering the bound of 0.037 on 
$\epsilon_{\tau \tau}^{u(d)}$ \cite{Farzan:2017xzy}, we find $q'_u,q'_d<10^{-4} (m_{Z'}/ 400~{\rm MeV})$. Notice that despite this bound, $q'_u$ ($q'_d$) in this model can still be 5 (10) times larger  than 
the corresponding couplings in the kinetic mixing model, $e \epsilon q_u$ ($e \epsilon q_d$).  If $Z'$ does not couple to the neutrinos at the tree level, its coupling to the quarks can be as large as $O(0.1)$. Then, to cancel gauge anomalies, new chiral fermions  have to be added.

As mentioned above, $\nu_\mu$ produced in the pion decay will be in fact a linear combination of $|\tilde{\nu}_\mu \rangle= \sum_{i=1}^3 |\nu_i\rangle \langle \nu_i |\nu_\mu \rangle$ for which $g_{\nu N}= g' U_{\mu 4}$ where $U_{\mu 4}$ is the mixing between $\nu_\mu$ and $N'$ (not with the heavier sterile particle, $N$)
so $g_{\nu N}$ can be as large as $10^{-2}$. On the other hand, the decay of heavier mesons such as $K^+$ or $D^+$ can produce both $|\nu_4 \rangle$ and $|\tilde{\nu}_\mu\rangle$. The flux from the heavy meson decay can be decomposed as a flux of $\nu_4$ proportional to $|U_{\mu 4}|^2$ plus a flux of $\tilde{\nu}_\mu$ proportional to $(1-|U_{\mu 4}|^2)$. While the scattering rate of $\nu_4 +{\rm quark}\to N+ {\rm quark}$ is proportional to $g'^2 (1-|U_{\mu 4}|^2)$, the scattering rate of $\tilde{\nu}_\mu +{\rm quark}\to N+ {\rm quark}$ should be proportional to $g'^2 |U_{\mu 4}|^2$. Let us denote the $\nu_\mu$ flux predicted (in the absence of new physics) from the pion decay and the heavy meson decay respectively by $F_\pi^\nu$ and $F_H^\nu$. The flux of $N$ from the interaction of $\nu_\mu$ within the toy model described in the appendix should then be given by $g_{\nu N}^2 (F_\pi^\nu +2F_H^\nu)$. At FASER$\nu$ for the $\nu_\mu$ energy lower than 1~TeV, we expect $F_\pi^\nu \gg F_H^\nu$ \cite{Abreu:2019yak} so we will neglect the contribution from $F_H^\nu$ in our analysis. For $\nu_\mu$ energy higher than 1~TeV, the flux from the Kaon decay dominates, $F_\pi^\nu \ll F_H^\nu$ \cite{Abreu:2019yak} so we will neglect $F_\pi^\nu$, taking into account the extra factor of 2 in front of $F_H^\nu$.

\section{Decay of new particles and their signatures at FASER$\nu$ \label{decay}}

In this section, we first discuss the decay products of $N$ assuming $Z'$ is kinetically mixed with
the photon. In the end, we shall comment on the case that $Z'$ is the gauge boson of the $B-3L_\tau$ symmetry.

\textbf{The $m_{Z'}>m_N$ option:} In the case $m_{Z'}>m_N$, the decay of $N$ will be three-body with decay rate
\be \Gamma(N \to N' e^- e^+)=\frac{g'^2 (e \epsilon)^2}{ \pi^3}\frac{m_N^5}{m_{Z'}^4}\left(\frac{1}{96}+\frac{13}{960} \left( \frac{m_{N'}^2}{m_{Z'}^2}\right)+O( \left( \frac{m_{N'}^2}{m_{Z'}^2}\right)^2)\right).\ee
For $m_N>m_{N'}+2 m_\mu,m_{N'}+2 m_\pi$, the decay modes $N\to N'\mu^-\mu^+, N'\pi^-\pi^+$ also open up. In the absence of a faster decay mode for $N$ (such as $N \to N' a\bar{a}$ as discussed in the appendix),
an $N$ particle with an energy of $E_N$ can decay after traveling a distance of
$$\Gamma^{-1}\frac{E_N}{m_N}\sim 0.2~{\rm m} \left(\frac{7\times 10^{-4}}{\epsilon}\right)^2\frac{1}{g'^2}\left(\frac{m_{Z'}}{\rm GeV}\right)^4 \left(\frac{0.5 ~{\rm GeV}}{m_N}\right)^6 \left( \frac{E_N}{500~{\rm GeV}}\right) .$$
Considering that $0.2$ m is smaller than the length of the detector, the majority of the decays take place within the detector in this parameter range. Since $N$ is highly boosted in the beam direction, its track will be aligned with the beam direction so the signature will be a shower similar to that expected in the neutral current interactions of neutrinos and a pair of $e^-e^+$ from a vertex separated from the shower vertex by $\sim 0.1$ m. Since $N$ is highly boosted in the forward direction,
the line connecting the two vertices should make an angle of $O(m_N/E_N)$ or smaller with the beamline direction.
The probability for one out of $N_{NC}\sim 5000$ background (SM) neutral current vertices being found behind the lepton vertex within a cone with an opening angle of $m_N/E_N$ is less than
$(\pi/3) (m_N/E_N)^2 (1.3/0.25)^2 N_{NC}$
which will be less than one. That is despite the separation, we can identify which of the observed NC vertices is associated with a certain observed dilepton.  The background neutral particles induced by photo-nuclear  interactions of muons \cite{Abreu:2019yak} can however complicate the analysis  of the data. Moreover, for the low energy-momentum transfer regime, the jets produced in the NC vertex may be too soft to be discernible. A complete analysis is beyond the scope of the present paper. If the kinematics allows, along the $e^-e^+$ pair signal there will be also signals of $\mu^-\mu^+$ or $\pi^+\pi^-$ emitted close to the direction of the beam.

The $e^- e^+$ pair can also come from the pair production by photons from the $\pi^0$ decay in the shower of neutral current interactions. Taking the cross-section of the pair production of $\sim 10~ b/$atoms \cite{34.15}, we find that the photons travel $\sim 1$ cm before pair production. Moreover, while the invariant mass of $e^-e^+$ from the photon will be of order of $2m_e$, that from
$N\to N' e^-e^+$ will be of $\sim m_N-m_{N'}\gg 2m_e$. Thus, by putting cuts on the distance between the shower vertex and the $e^-e^+$ vertex and the invariant mass of $e^-e^+$, the background can be substantially reduced. Considering that the total neutral current events for FASER$\nu$ are only O(5000), applying these cuts the number of background from $\pi^0$ should become negligible. Photons can be also induced by photo-nuclear interactions of muons (see Table IV of \cite{Abreu:2019yak}). Again by applying cuts on the invariant mass of the final charged leptons, the background from pair production can be eliminated.
Similarly, but with a rate suppressed by $m_e^2/m_\mu^2$, the photons can produce muon pair \cite{cern}. The invariant mass of the muon antimuon pair from the photon will be close to $2m_\mu$.
As long as $m_N-m_{N'}\gg 2 m_\mu$, again by the measurement of the invariant mass of the charged lepton pair, the background can be vetoed. Another source of background for the $\mu^-\mu^+$ pair is
$\nu_\mu+{\rm nucleon}\to \mu+c+X$ and the subsequent decay of $c$ into $\mu$. For contained events, this background can be reduced by the event topology
({\it i.e.,} reconstructing the $D$ meson track).
Similar consideration applies for the background from $\nu_e +{\rm nucleon} \to e+c+X$ and the subsequent decay $c\to e +X$ to the $e^-e^+$ signal.

By measuring the distance between the neutral current vertex and the $N$ decay vertex, one can derive some information on the $N$ lifetime. However, since one of the final particles ($N'$) will be missing, the derivation of the energy of $N$ and hence the boost factor will not be possible. The derivation of the lifetime will also suffer from low statistics.

As discussed in sect. \ref{model} and in the appendix, it is possible to obtain four muon signal through $N \to N'Z^{' (*)}\to N' a \bar{a}$ and subsequently $a\to \mu \bar{\mu}$ and $\bar{a}\to \mu \bar{\mu}$. In order for $N \to N' Z^{' (*)}\to N' a \bar{a}$ to dominate over $N\to N' Z^{' (*)}\to N' e^-e^+$, both $N$ and $Z'$ should be heavier than $2 m_a$ (and hence heavier than $4m_\mu$) and moreover, the coupling of $a$ to $Z'$ should be much larger than $e\epsilon$ which means $N$ will decay promptly but the decay length of $a$ can be within the range $1~{\rm mm}-$few meters. Again, the $a$ and $\bar{a}$ particles will be highly boosted in the forward direction so the muon pairs will be emitted along the beam.
If the event is fully contained in the FASER$\nu$ detector by measuring the invariant mass of $\mu^-\mu^+$ pair, the mass of $a$ particle can in principle be derived. Moreover, by reconstructing the momenta of $\mu$ and $\bar{\mu}$, the directions of $a$ and $\bar{a}$ and therefore the $N$ decay vertex can be deduced. The distance between the $N$ decay vertex and the $a$ and $\bar{a}$ decay vertices is a measure of the $a$ and $\bar{a}$ lifetime. However, for $q'_q=e\epsilon q_q$ smaller than the BABAR bound, the statistics of the contained event will be too low to perform such analysis. As discussed in the previous section, the four muon events originated in the rock before the detector can increase the statistics. We will discuss this possibility further in sect \ref{production}.

{\bf{ The $m_{Z'}<m_N$ option:}} If $m_N>m_{Z'}$, $N$ can immediately decay into $Z'N'$ and $Z'\nu_\mu$. $Z'$ can then go through a two body decay with
$$\Gamma(Z'\to f \bar{f})=\frac{(e\epsilon q_f)^2}{12\pi}m_{Z'}\left(1+2\frac{m_f^2}{m_{Z'}^2}\right)
\left(1-4\frac{m_f^2}{m_{Z'}^2}\right)^{1/2},$$
where $f$ can be $e$ or $\mu$ and the factor of 3 in the denominator comes from averaging over three polarizations of the initial vector boson. $Z'$ can also decay into $\pi^+\pi^-$ but we shall focus only on the leptonic decay modes. The decay length of $Z'$ is of order of $0.1~{\rm cm}(0.1~{\rm GeV}/m_{Z'})^2 (E_{Z'}/250~{\rm GeV})$.
If $m_{Z'}>200~$MeV, the decay length will be shorter than 1 mm, so FASER$\nu$ cannot disentangle the track of $Z'$. In this case, $Z'$ can decay into $\mu^+\mu^-$ so if the $\nu_\mu$ scattering takes place inside the detector, the signal will be a nuclear shower and a $\mu^-\mu^+$ pair from a single vertex which is background free. By measuring the energy momentum of $\mu^+\mu^-$, $m_{Z'}$ can be reconstructed. Another signal will be a nuclear shower plus $e^-e^+$ pair. Again the invariant mass of the $e^-e^+$ pair gives $m_{Z'}$. Since $Z'$ will be highly boosted, the angle between the final fermion pair will be small and of order of $m_{Z'}/E_{Z'}\sim 10^{-3}$ but the angular resolution of FASER$\nu$ is better than 0.1 milliradian \cite{Abreu:2019yak} so this angle can be reconstructed. We can write
$$m_{Z'}^2=2m_f^2+2\left[\sqrt{p_f^2+m_f^2}\sqrt{p_{\bar f}^2+m_{\bar f}^2}-p_fp_{\bar{f}}\cos\theta\right]$$ where $\theta$ is the angle between $f$ and $\bar{f}$. Since there will be an uncertainty of $30\%$ in the reconstruction of $p_f$ and $p_{\bar{f}}$ ({\it i.e.,} $\delta p_f/p_f\sim 30 \%$ \cite{Abreu:2019yak}), the uncertainty of $m_{Z'}$ derived from a single pair of $f \bar{f}$ will be 40\%. Of course, if the statistics is high, the uncertainty in the derivation of $m_{Z'}$ will be reduced. If $N_{pair}$ pairs are registered, the uncertainties will be
$40 \%/\sqrt{N_{pair}}$.
As discussed in the appendix and above, by coupling the $Z'$ particle to a pair of $a$ scalars lighter than $m_{Z'}/2$, we can have a dominant four-muon signal which is background free. This of course requires $Z'$ to be heavier than $4m_\mu$.

Notice that in both cases $m_{Z'}>m_N$ and $m_{N}>m_{Z'}$, the main decay mode of $N$ produces $N'$ which is a metastable particle. As discussed in sect. \ref{model}, the lifetime of $N'$ will be long enough to exit the detector so $N'$ will appear as missing energy-momentum.

As mentioned in sect. \ref{Enhance}, the neutrinos can interact in the soil and concrete before reaching the detector. In this case, the $e^-e^+$ or $\pi^-\pi^+$ pairs (as well as the neutral current showers) will be absorbed but the muon pairs can reach the detector. Since the angular deviation of muon and antimuon during the propagation will be much larger than the angle that they make with each other at the production ($m_{Z'}/E_{Z'}\sim 10^{-3}$), these through-going muons cannot be used to derive the mass of $Z'$; however, they will be indicative of new physics and taking them into account a stronger bound on the coupling can be obtained. We shall study this in more detail.

Let us now briefly discuss the case that $Z'$ is the gauge boson of $B-3 L_\tau$. If the kinematics allows $Z'$ can decay into $\pi^+ \pi^-$ or $\tau^+ \tau^-$ but as discussed before we are interested in the case that the decay mode $N\to N'Z^{'(*)} \to N' a \bar{a}$ dominates and we obtain a four muon signature. The rest of the discussion is similar to above, except that here the BABAR bounds do not apply so the statistics of the contained vertex events can be larger. By measurement of the invariant mass of the contained vertex muon pairs, the mass of $a$ can also be derived with a precision of $40 \% /\sqrt{N_{pair}}$ where $N_{pair}$ is the number of contained vertex muon pair. For $m_{Z'}<m_N$, the invariant mass of $\mu \bar{\mu}\mu \bar{\mu}$ gives the mass of $m_{Z'}$ with a precision of $(\delta E_\mu/E_\mu)\sqrt{4/N_{pair}}=60 \% /\sqrt{N_{pair}}$. For $m_{Z'}>m_N$, the invariant mass of $\mu \bar{\mu}\mu \bar{\mu}$ will have a continuous spectrum and will not be peaked at $m_{Z'}$ but $m_a$ can still be derived by extracting the invariant mass of the $\mu \bar{\mu}$ pairs.

\section{Production of $N$ via neutrino flux and the signal at FASER$\nu$ \label{production}}

In sect. \ref{DIS}, we  discuss the $N$ production rate by scattering of neutrino beam off the nuclei in the deep inelastic scattering regime, assuming $Z'$ is kinetically mixed with
the photon. The same discussion holds valid for the case that $Z'$ is the gauge boson of the $B-3L_\tau$ symmetry, replacing $e\epsilon q_q$ with the common gauge coupling of the quarks. We then discuss the number of signal produced by scattering  both inside the detector and in the rock before the detector.
In sect. \ref{Coh-regime},
we  show that for relatively light $Z'$ coherent scattering can dominate over deep inelastic scattering and discuss the production rates taking into account enhancement due to coherence.  We discuss the difference in the form of signal in this regime and that in the deep inelastic regime. In sect. \ref{electron-regime}, we study the $N$ production by scattering of the neutrino beam off the electrons and show that it is negligible. 

\subsection{Deep Inelastic Scattering regime\label{DIS}}
The cross section of the scattering of $\nu_\mu\simeq \tilde{\nu}_\mu=\sum_{i=1}^3 U_{\mu i}\nu_i$ off quarks can be written as
\be \frac{d\sigma (\nu_\mu +q \to N+q)}{d\cos \theta} \simeq \frac{d\sigma (\tilde{\nu}_\mu +q \to {N}+q)}{d\cos \theta}= \frac{g_{\nu N}^2 (e q_q \epsilon)^2}{32 \pi} \frac{(s-m_N^2)^2}{s^2}\frac{5s+m_N^2+2s\cos\theta +(s-m_N^2)\cos^2\theta}{\left( (s-m_N^2) (1-\cos \theta)+m_{Z'}^2\right)^2} ~,\label{diff-sig}\ee
where $g_{\nu N}=g'U_{\mu 4}$ and $s$ is the Mandelstam variable for the system of $\nu_\mu$ and parton carrying a fraction $x$ of the proton momentum, $s=2 x m_p E_{\nu_\mu}\sim (10~{\rm GeV})^2(x/0.1)(E_{\nu_\mu}/500~{\rm GeV})$
and $\theta$ is the scattering angle in the center of mass frame of the $\nu_\mu$-parton system. $q_q$ is the electric charge of the parton; {\it i.e.,} $q_u=2/3$ and $q_d=-1/3$. Neglecting $m_{N'}^2/s$, we can write $\sigma(N'+q \to N+q)=\sigma(\nu_\mu +q \to N+q)(g'/g_{\nu N})^2$ in which $(g'/g_{\nu N})^2=U_{\mu 4}^{-2}$.

To obtain the $\nu_\mu$ cross section off the nucleus, we should convolute with the Parton Distribution Functions (PDFs)
\be \frac{d \sigma_{\nu N}^{tot}(E_{\nu_\mu})}{d\cos \theta}=\sum_{q \in \{u,d,s\}}\int_{m_N^2/(2m_pE_{\nu_\mu})}^1 [F_q(x,t)+F_{\bar{q}}(x,t)] \frac{d\sigma (\nu_\mu +q \to N+q)}{d\cos \theta}dx ~, \label{Fqxt} \ee
where $F_q$ and $F_{\bar{q}}$ are parton distribution functions and $t=(m_N^2-s)(1-\cos\theta)$.
Notice that as $\theta \to 0$, $t$ goes to zero. This can be understood as we have neglected the masses of the partons and the $t$ variable associated with the scattering of a massless particle to another in the forward direction vanishes. In the limit $t\to 0$,
$d\sigma/d\cos \theta$ in Eq. (\ref{diff-sig}) will be dramatically enhanced. In fact, the total cross-section is
\be \sigma (\nu_\mu +q \to N+q)=\int_{-1}^{+1} \frac{d\sigma (\nu_\mu +q \to N+q)}{d\cos \theta} d\cos\theta=\label{sigg} \ee $$\frac{g^2_{\nu N}(eq_q\epsilon)^2}{16\pi s^2}\left(\left(-m_N^2+m_{Z'}^2+2 s\right) \log \left(\frac{m_{Z'}^2}{-2 m_N^2+m_{Z'}^2+2 s}\right)+\frac{2 \left(s-m_N^2\right) \left(-2 m_N^2 \left(m_{Z'}^2+2 s\right)+m_{Z'}^4+3
m_{Z'}^2 s+4 s^2\right)}{m_{Z'}^2 \left(-2 m_N^2+m_{Z'}^2+2 s\right)}\right) ~,$$
in which $s=2xm_p E_{\nu_\mu}$. The $1/m_{Z'}^{2}$ behavior of $ \sigma (\nu_\mu +q \to N+q)$ shows that for the majority of the scatterings, $|t|\sim m_{Z'}^2$ or smaller. On the other hand, the dependence of PDFs on $t$ is mild. We can therefore simplify the integration in Eq. (\ref{Fqxt}) by setting
$F_q(x,t)=F_q(x,-m_{Z'}^2)$ and $F_{\bar{q}}(x,t)=F_{\bar{q}}(x,-m_{Z'}^2)$ and write
$$\sigma_{\nu N}^{tot}(E_{\nu_\mu})= \int \frac{d \sigma_{\nu N}^{tot}(E_{\nu_\mu})}{d\cos \theta}d\cos \theta=\sum_{q \in \{ u, d,s \}}
\int_{\frac{m_N^2}{2m_pE_{\nu_\mu}}}^1 [F_q(x,-m_{Z'}^2)+F_{\bar{q}}(x,-m_{Z'}^2)]\sigma (\nu_\mu +q(x) \to N+q) dx.$$

Notice that we have used the fact that the cross sections of the scattering off quark and antiquarks are equal. Similarly, the scattering cross sections of $\nu_\mu$ and $\bar{\nu}_\mu$ off quarks are equal. The number of $\nu_\mu$ and $\bar\nu_\mu$ converting to $N$ inside FASER$\nu$ can therefore be estimated as
\be N_{HESE}=\frac{M_{det}}{m_p} \int \sigma_{\nu N}^{tot}(E_\nu) [F_{\nu_\mu}(E_\nu)+F_{\bar{\nu}_\mu}(E_\nu)][1+\Theta (E_\nu-1~{\rm TeV})] dE_\nu, \label{HESE}\ee
$M_{det}$ is the total mass of FASER$\nu$. $F_{\nu_\mu}(E_\mu)$ and $F_{\bar\nu_\mu}(E_\mu)$ are the fluxes of $\nu_\mu$ and $\bar\nu_\mu$ at the detector (integrated over time). $F_{\nu_\mu}(E_\mu)$ and $F_{\bar\nu_\mu}(E_\mu)$ can be read from Fig. 4 of \cite{Abreu:2019yak}.
For $E_\nu<$TeV, the majority of the neutrino flux comes from the
pion decay. Since pion is lighter than $m_\mu+m_{N'}$, the flux will be purely composed of $\tilde{\nu}_\mu \simeq \sum_i U_{\mu i}\nu_i$. For $E_\nu>$TeV, the Kaon decay dominates so the flux at the detector can be decomposed as the $\tilde{\nu}_\mu$
part which takes a fraction of $(1-|U_{\mu 4}|^2)$ of the flux plus the $N'$ part which constitutes a fraction of $|U_{\mu 4}|^2$ of it. Since the cross section of $N'$ is $|U_{\mu 4}|^{-2}$ times that of $\tilde{\nu}_\mu$, for $E_\nu>$TeV, the contribution from $N'$ to $N_{HESE}$ should be equal to that from $\tilde{\nu}_\mu$. The $\Theta$-function in Eq. (\ref{HESE}) accounts for the contribution from $N'$.


Let us now study how much the discovery reach of FASER$\nu$ can be improved by including $\nu_\mu +{\rm nucleon} \to N +X$ taking place in the rock before the detector. As discussed before for the through-going events, we cannot discriminate the signal and background dimuon events. We therefore assume $m_N>m_{N'}+4 m_\mu$ and suppose the signal is the background-free $\mu \bar{\mu}\mu \bar{\mu}$ as described in the appendix. As mentioned before, there are 10 meters of concrete and 90 meters of rock before the detector. As demonstrated in the right panel of Fig. 5 of \cite{Abreu:2019yak}, the $\nu_\mu$ beam reaching the FASER$\nu$ detector is very well collimated along the proton beamline. For simplicity, we shall assume that all $\nu_\mu$ flux
is directed towards the center of the detector.
At the scattering of $\nu_\mu$ off a parton, the angle in the lab frame $\theta_{lab}$ (the angle between the produced $N$ and the initial $\nu_\mu$) will be smaller than $m_{Z'}/E_\nu<10^{-3}$. The angular spread of the muon particle from the direction of $N$ can be estimated as $ \theta'_{lab}\sim (m_N -3 m_\mu)/(2E_N) < {\rm few }\times 10^{-3}$.
As mentioned before the angular dispersion due to propagation of the produced muons is of order of $0.5^\circ=0.0087$ radians \cite{Antonioli:1997qw} which is larger than $ \theta_{lab}$ and $ \theta'_{lab}$.
As we discussed before, in order to produce two muon pair signals ($\mu \bar{\mu}\mu \bar{\mu}$), the $Z'$ particle has to have a large coupling to an intermediate scalar, $a$. Thus, the whole $N\to N' Z^{'(*)}\to N' a\bar{a}$ process will take place promptly. However, the $a$ particles can travel a sizable distance before producing $\mu \bar{\mu}$.
If the production of the $\mu \bar{\mu}$ pair takes place at a distance smaller than $(0.25~{\rm m}/2)/0.0087=14~{\rm m}$, practically both $\mu^-$ and $\mu^+$ from the $a$ decay will reach the detector.
If the production takes place farther, we should multiply the rate with the probability that both $\mu^-$ and $\mu^+$ reach the detector. If the production takes place at a distance $z>14$ m from the detector, the probability of each of $\mu^-$ or $\mu^+$ arriving at the detector will be
\be p_\mu(z) =\frac{(0.25~{\rm m})^2}{4\pi z^2}\times \frac{2}{(0.0087)^2} \ee so the probability for both muons reaching the detector will be given by $p_\mu^2$. Remember that if only one of them reaches the detector, it cannot be distinguished from the background muons originating from the IP.
In fact, both pairs from the $a$ and $\bar{a}$ decay have to pass through the detector in order to distinguish the signal event from the background. Let us suppose $a$ and $\bar{a}$ respectively decay at distances of $z_1$ and $z_2$ from the detector. The probability that all the four muon and antimuons arrive at the detector is given by $[p(z_1)]^2\times [p(z_2)]^2$.

As long as the distance traveled by $a$ and $\bar{a}$ before decay is shorter than 10~m, the number of muon pairs reaching the detector originated from the en-route rock and concrete can be estimated as
\be N_{through-going}= \int \sigma_{\nu N}^{tot}(E_\nu) [F_{\nu_\mu}(E_{\nu})+F_{\bar\nu_\mu}(E_{\nu})] [1+\Theta (E_\nu-1~{\rm TeV})] dE_\nu \times \label{8} \ee $$\left[ \frac{\rho_{con}}{m_p}\int_{90~{\rm m}}^{100~{\rm m}} [p_\mu(z)]^4 dz +\frac{\rho_{soil}}{m_p}\int_{14~{\rm m}}^{90~{\rm m}} [p_\mu(z)]^4 dz +\frac{\rho_{soil}}{m_p}(14~{\rm m}) \right],$$
where we have assumed that practically all $N$ decays lead to a $\mu \bar{\mu}\mu \bar{\mu}$. $\rho_{soil}$ and $\rho_{con}$ are respectively the soil and concrete densities. To carry out our computation, we take $\rho_{soil}=\rho_{con}=2.5$ gr/cm$^3$.
As discussed in sect \ref{Enhance}, if FASER$\nu$ is equipped with the scintillator plate in its front and records the timing of the arrival of the incoming $\mu \bar{\mu}\mu \bar{\mu}$ event, the background will be negligible. As a result, even the detection of a single $\mu \bar{\mu}\mu \bar{\mu}$ event can be regarded as an indication for new physics.
\subsection{Coherence regime\label{Coh-regime}}
As discussed after Eq. (\ref{sigg}),  in the majority of the scatterings for light $Z'$, the energy-momentum transfer, $\sqrt{t}\sim m_{Z'}\ll \sqrt{s}$. As a result for $m_{Z'}<100$ MeV, the scattering  on the whole nucleus can be coherent.  Then, the scattering cross section off a nucleus with a  mass number of $A$, a mass of $m_A$ and an atomic number of $Z_A$ can be written as 

\be \frac{d\sigma (\nu_\mu +A \to N+A)}{d\cos \theta} \simeq \frac{d\sigma (\tilde{\nu}_\mu +A \to {N}+A)}{d\cos \theta}=
\label{coh-scattering} \ee $$ |F(t)|^2\times \frac{g_{\nu N}^2 (e Z_A\epsilon)^2}{32 \pi} \frac{(s-m_N^2)^2}{s^2}\frac{5s+m_N^2+2s\cos\theta +(s-m_N^2)\cos^2\theta}{\left( (s-m_N^2) (1-\cos \theta)+m_{Z'}^2\right)^2} ~,$$
where $F(t)$ is the form factor. Notice that this formula is similar to Eq. (\ref{diff-sig}) except that $q_q$ is replaced by $Z_A$. Instead of the dark photon (kinetic mixing) model, if we take the model of  $B-3L_\tau$ or $B$ gauge symmetry, we have to replace $e Z_A\epsilon$ with $3q'_u A$. The $s$ Mandelstam variable here is equal to $m_A^2+2E_\nu m_A$ where $E_\nu$ is the energy of the incoming neutrino in the lab frame. The cross section of anti-neutrinos is again given by Eq. (\ref{coh-scattering}).

For $F(t)$, we may take the Helm form factor \cite{Helm} as
$$ |F(t)|^2=\left( \frac{3 j_1(R_1 \sqrt{t})}{R_1 \sqrt{t}}\right)^2 \exp^{-t s_1^2}$$
where $j_1(x)=\sin x/x^2-\cos x/x$ and $R_1=(c^2+7\pi^2a^2/3-5s_1^2)^{1/2}$ in which $s_1 =0.9$ fm, $a_1=0.52$ fm and $c=1.23 A^{1/3}-0.6$ fm. The detector is made of Tungsten with $Z_A=74$, $m_A=138$ GeV and $A=138$. Soil is mostly composed of  Silicon with $Z_A=14$, $m_A=28$ GeV and $A=28$.

For $\sqrt{t}\sim m_{Z'}<50$ MeV (100~MeV), $|F(t)|^2$ is larger than 0.5 (than $10^{-2}\sim 1/Z_W$) but for higher $\sqrt{t}$,  $|F(t)|^2$ exponentially drops, reflecting the fact that for large $\sqrt{t}$, the coherence is destroyed. For $ 100~{\rm MeV}<\sqrt{t}<500$~MeV, the scattering off nucleus will be neither coherent nor in the deep inelastic scattering regime. For $\sqrt{t}\sim 100-500$ MeV, the scattering can be treated as quasi-elastic. Exploring the whole range is not the goal of this paper, we therefore focus on two coherent and deep inelastic scattering regimes.

Notice that when scattering is coherent, we shall not have the NC jets and the signal will be only composed of  charged leptons from the $N$ decay without any discernible NC vertex. The produced $N$ will have an energy close to the initial neutrino so the final leptons  will be energetic enough to be detected.

The rest of the discussion is similar to what we had in sect.
\ref{DIS}, except that here $\rho/m_p$ should be replaced by
$\rho/m_A$ in Eq. (\ref{HESE}).
\subsection{Scattering off the electrons\label{electron-regime}}
In the $B-3L_\tau$ or $B$ gauge model, $Z'$ does not couple to the electron so the production of $N$ by the scattering off electrons is negligible. However, in the dark photon model, the cross section  $d\sigma (\nu_\mu+e \to N+e)/d\cos\theta =
d\sigma (\bar{\nu}_\mu+e \to \bar{N}+e)/d\cos\theta $ is given again by Eq. (\ref{sigg}), replacing $q_q$ with 1. The Mandelstam variable is $s\simeq 2m_e E_\nu= {\rm GeV}^2 (E_\nu /{\rm TeV})$.

	If the scattering takes place inside the detector, signal will be composed of  a boosted electron plus leptons from $N$ decay. For the scatterings inside the rock, the boosted electron will be trapped and cannot reach the detector. The number density of electrons in Tungsten is given by $(74/183)\rho_W/m_p$ and that in soil is approximately equal to $\rho_{soil}/(2m_p)$.
	Similarly to the SM scattering of neutrinos off matter, the cross section of scattering off the electron is negligible in comparison to that of scattering off the nuclei.  
	This is understable  both in the Deep Inelastic Scattering (DIS) regime and in the coherent regime: In the DIS regime, the system of neutrino parton with $x>10^{-3}$ has larger $s$ compared to that of the  neutrino electron system so the cross section is larger. For lower values of $x$, the scattering off nucleus receives a large enhancement from the large number of target sea quarks. For light $Z'$, the scattering off nuclei receives enhancement due to the coherence. We therefore neglect scattering off the electron in our analysis which has a cross section suppressed by one or two orders of magnitude compared to the cross section of scattering off nucleons for relatively light $N$ production.
	 Moreover, the energy of center of mass of the neutrino electron system will be too small to lead to the production of $N$ with a mass larger than 2 GeV.  
 \section{Results \label{res}}
 
 \begin{figure}[h]
\hspace{0cm}
\includegraphics[width=0.6\textwidth, height=0.45\textwidth]{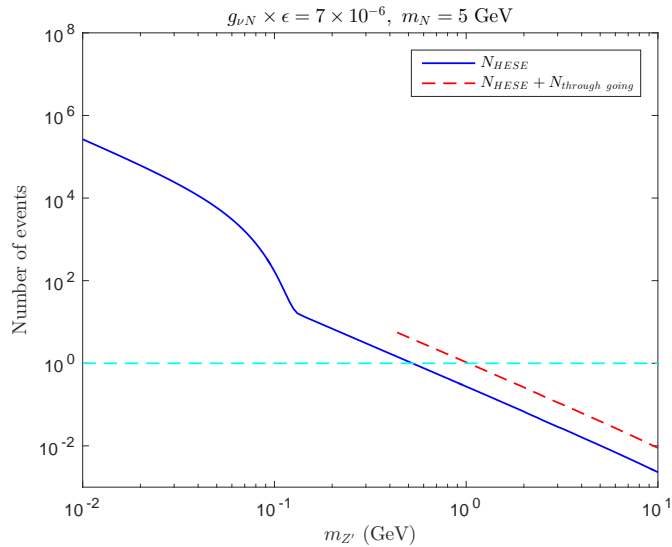}
\caption[...]{
Number of new physics events at FASER$\nu$ during LHC run III vs. $m_{Z^\prime}$. The solid blue line shows the number of $N$ produced inside the detector ($N_{HESE}$). The red dashed line shows the total number of the $\mu \bar{\mu}\mu \bar{\mu}$ events coming from
the production of $N$ both inside the detector and en-route rock and concrete ($N_{HESE}+N_{through-going}$). The line is clipped at $m_{Z'}=4m_\mu$ because below this value, the $\mu \bar{\mu}\mu \bar{\mu}$ signal cannot be created. We have set $m_N=5~\rm{GeV}$ and $g_{\nu N}\times\epsilon=7\times10^{-6}$ and have taken the flux of neutrinos from Fig. 4 of \cite{Abreu:2019yak}.
The cyan horizontal dashed line corresponds to the number of events equal to one.
\label{NOEmZp}
}
\end{figure}

\begin{figure}[h]
\hspace{0cm}
\includegraphics[width=0.6\textwidth, height=0.45\textwidth]{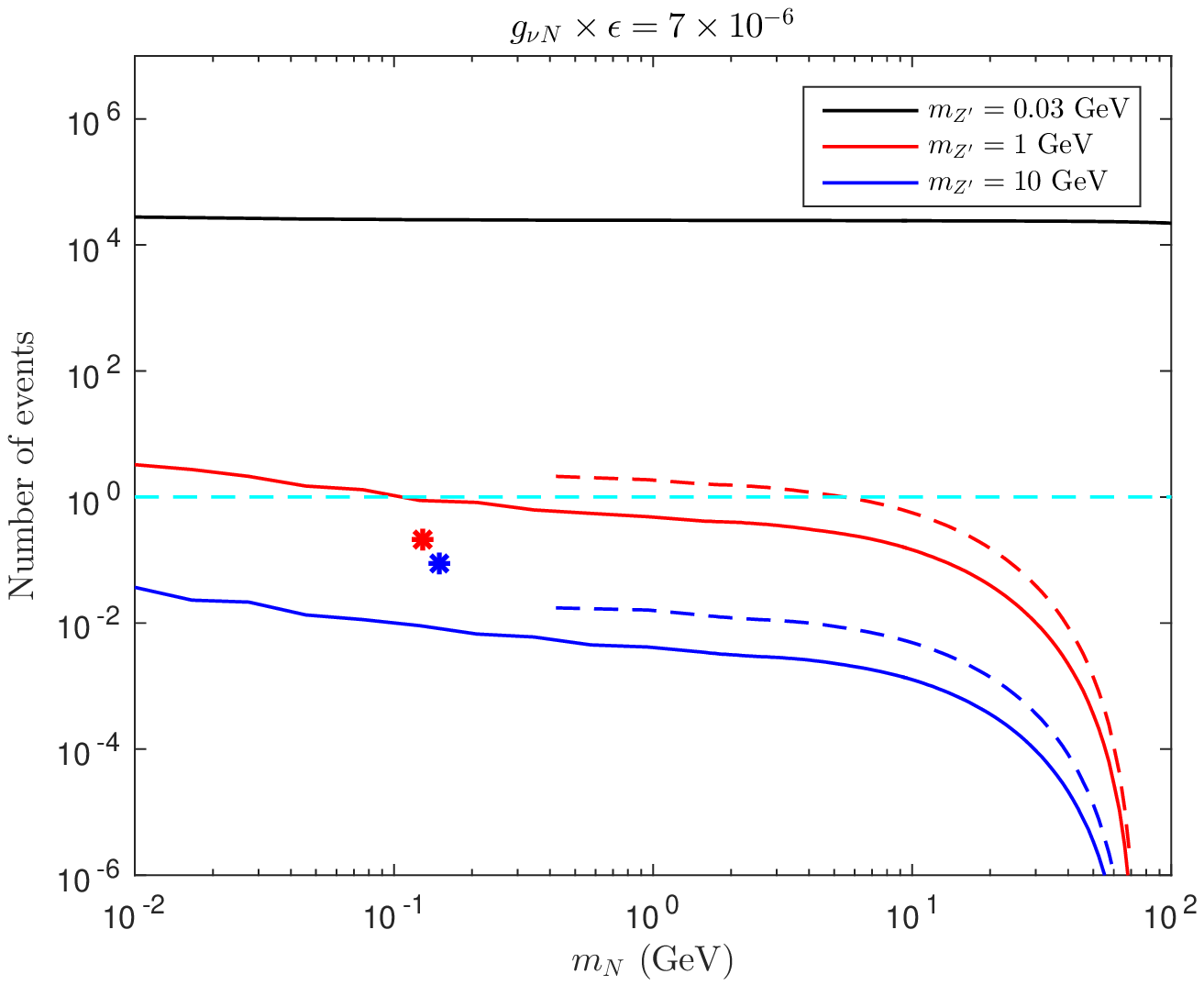}
\caption[...]{
Number of new physics events at FASER$\nu$ during LHC run III vs. $m_{N}$. The solid curves show the number of $N$ produced inside the detector ($N_{HESE}$). The dashed curves show the total number of the $\mu \bar{\mu}\mu \bar{\mu}$ events coming from production of $N$ both inside the detector and en-route rock and concrete ($N_{HESE}+N_{through-going}$). The
dashed curves are clipped at $m_N=4m_\mu$ as for lighter $N$, the signal of $\mu \bar{\mu}\mu \bar{\mu}$ for through-going muons cannot be obtained.
We have set $g_{\nu N}\times\epsilon=7\times10^{-6}$ and have taken the flux of neutrinos from Fig. 4 of \cite{Abreu:2019yak}. The black, red and blue curves respectively correspond to $m_{Z^\prime}=0.03,~1$ and 10 $\rm{GeV}$. The blue and red dots show the number of events corresponding to the best fit solution of MiniBooNE taken from Ref.~\cite{Ballett:2018ynz} with $\epsilon g_{\nu N}=2.7\times 10^{-6}$, $m_N=150$ MeV, $m_{Z'}=1.2$ GeV and from Ref.~\cite{Bertuzzo:2018itn} with $\epsilon g_{\nu N}=2\times 10^{-8}$, $m_N=130$ MeV, $m_{Z'}=30$ MeV, respectively. The cyan horizontal dashed line corresponds to the number of events equal to one.
\label{NOEmN}
}
\end{figure}
Let us now study the number of events at FASER$\nu$ and then discuss what bounds can be derived from this experiment. For the time integrated neutrino fluxes, we have used Fig 4 of \cite{Abreu:2019yak} which corresponds to the run III of the LHC during 2022-2024. For Parton Distribution Function (PDF),
we have taken the CT10 model \cite{Lai:2010vv} from LHAPDF-6.3.0 software \cite{Buckley:2014ana}.

Fig.~\ref{NOEmZp} shows the number of produced $N$ versus $m_{Z'}$ for $20 ~{\rm MeV}<m_{Z'}<10$~GeV, taking $g_{\nu N}=10^{-2}$
and $\epsilon=7 \times 10^{-4}$. To draw this plot, we have set $m_N=5$ GeV so, as long as $m_{Z'}>4 m_\mu$ the kinematics allows the production of four muons from $N$ decay. As seen from
these figures, including the through-going events increases the statistics by a factor of 4. For
$m_{Z'}<1~{\rm GeV}~ (0.5~{\rm GeV})$, $N_{HESE}+N_{through-going}$ ($N_{HESE}$) will exceed 1 so hints for new physics will be revealed in Run III of the LHC. Notice that $m_N=5$ GeV is well beyond the reach of MINER$\nu$A and CHARM II experiments \cite{Arguelles:2018mtc}.
Below $m_{Z'}<0.1$ GeV, the main contribution comes from the coherent scattering.

Fig.~\ref{NOEmN} shows the number of produced $N$ versus $m_N$ for various values of $m_{Z'}$, again taking $g_{\nu N}=10^{-2}$
and $\epsilon=7 \times 10^{-4}$. As seen from this figure, if $m_{Z'}< 0.1$ GeV, thanks to the coherent enhancement, the number of the events at FASER$\nu$ can be as large as few times $10^4$. In this regime, the Mandelstam variable $s=m_A^2+2E_\nu m_A \gg m_N$ so, the number of events does not change with $m_N$ below $m_N\sim 100$ GeV. For $m_{Z'}>$few GeV, the number of events will be less than 1 so FASER$\nu$ during the run III cannot test
$Z'$ heavier than $\sim 2$ GeV but during the high luminosity run of the LHC and with an upgrade of FASER$\nu$, the parameter range with heavier values of $Z'$ can also be probed.
For $m_N<4 m_\mu$, the kinematics does not allow the background free $\mu\bar{\mu}\mu\bar{\mu}$ signal so the through-going events cannot help to extend the effective volume of FASER$\nu$.
For $m_N<2m_\mu$, the $\mu \bar{\mu}$ signal will also be absent but $e^-e^+$ can be produced and detected inside FASER$\nu$.

The blue and red dots in Fig.~\ref{NOEmN} correspond to the best fit of the MiniBooNE solutions, respectively, in Ref. \cite{Ballett:2018ynz} with $\epsilon g_{\nu N}=2.7\times 10^{-6}$, $m_N=150$ MeV, $m_{Z'}=1.2$ GeV and in Ref. \cite{Bertuzzo:2018itn} with $\epsilon g_{\nu N}=2\times 10^{-8}$, $m_N=130$ MeV, $m_{Z'}=30$ MeV. Notice that at MiniBooNE, the neutrino energy in the beam is much lower. Because of light $Z'$ \cite{Bertuzzo:2018itn}, the scattering off nucleons in the nucleus at MiniBooNE is coherent, leading to a huge enhancement of the cross-section by the square of the atomic number for a given coupling.  As seen from the figure, the number of events during run III of the LHC at FASER$\nu$ for the values of parameters providing the best fit   to MiniBooNE will be less than 1. With an upgrade of FASER$\nu$ during the high luminosity run of the LHC, the statistics can reach 200 times larger \cite{Abreu:2019yak}. These solutions can then be tested.

As shown in the figure even in the deep inelastic scattering regime with $m_{Z'}>100$ MeV, as $m_N$ varies between 0.1~GeV to 6 GeV, the number of events only slightly changes.
In general, for $s\gg m_N^2$, we expect only a weak dependence on $m_N$ but for $s\sim m_N^2$, the dependence should be strong so the weak dependence of the number of events on $m_N$ means the main contribution to $\sigma_{\nu N}^{tot}$ comes from relatively large values of $x$ for which $s\gg m_N^2$. As seen from Fig.~\ref{NOEmZp}, the dependence of the number of events on $m_{Z'}$ is however strong. In the limit $s\gg m_{Z'}^2$, from Eq. (\ref{sigg}), we find $\sigma \propto 1/m_{Z'}^2$ which receives the dominant contribution from small scattering angles; {\it i.e.,} $\theta \lsim m_{Z'}/\sqrt{s}$ in Eq. (\ref{diff-sig}) where $\theta$ is the scattering angle in the center of mass frame of the parton neutrino system. The corresponding scattering angle in the lab frame is $\theta_{lab}< \theta \sqrt{s}/E_\nu =m_{Z'}/E_\nu=10^{-3}(m_{Z'}/1 ~{\rm GeV})$.

\begin{figure}[h]
\hspace{0cm}
\includegraphics[width=0.493\textwidth, height=0.36\textwidth]{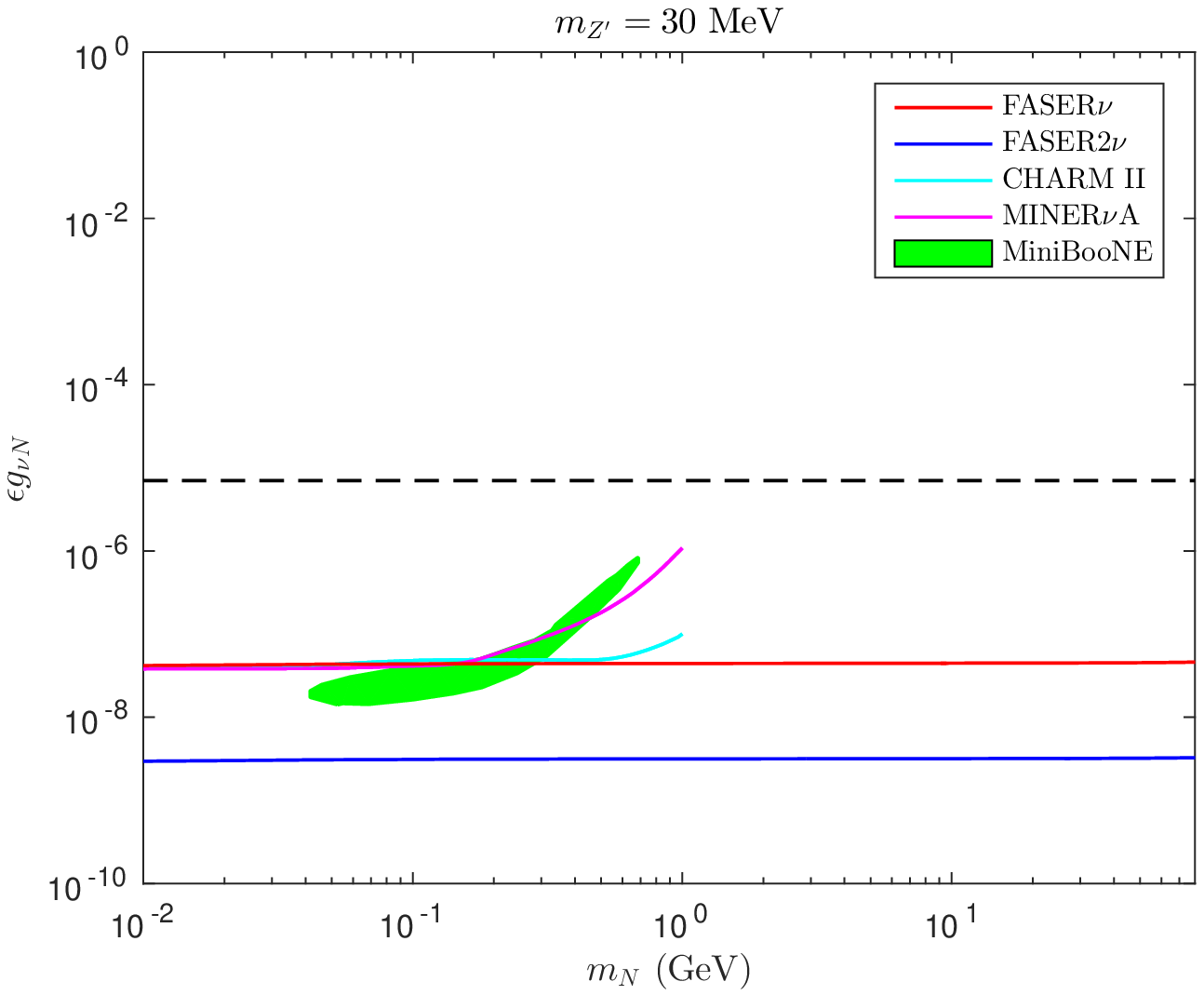}
\includegraphics[width=0.493\textwidth, height=0.36\textwidth]{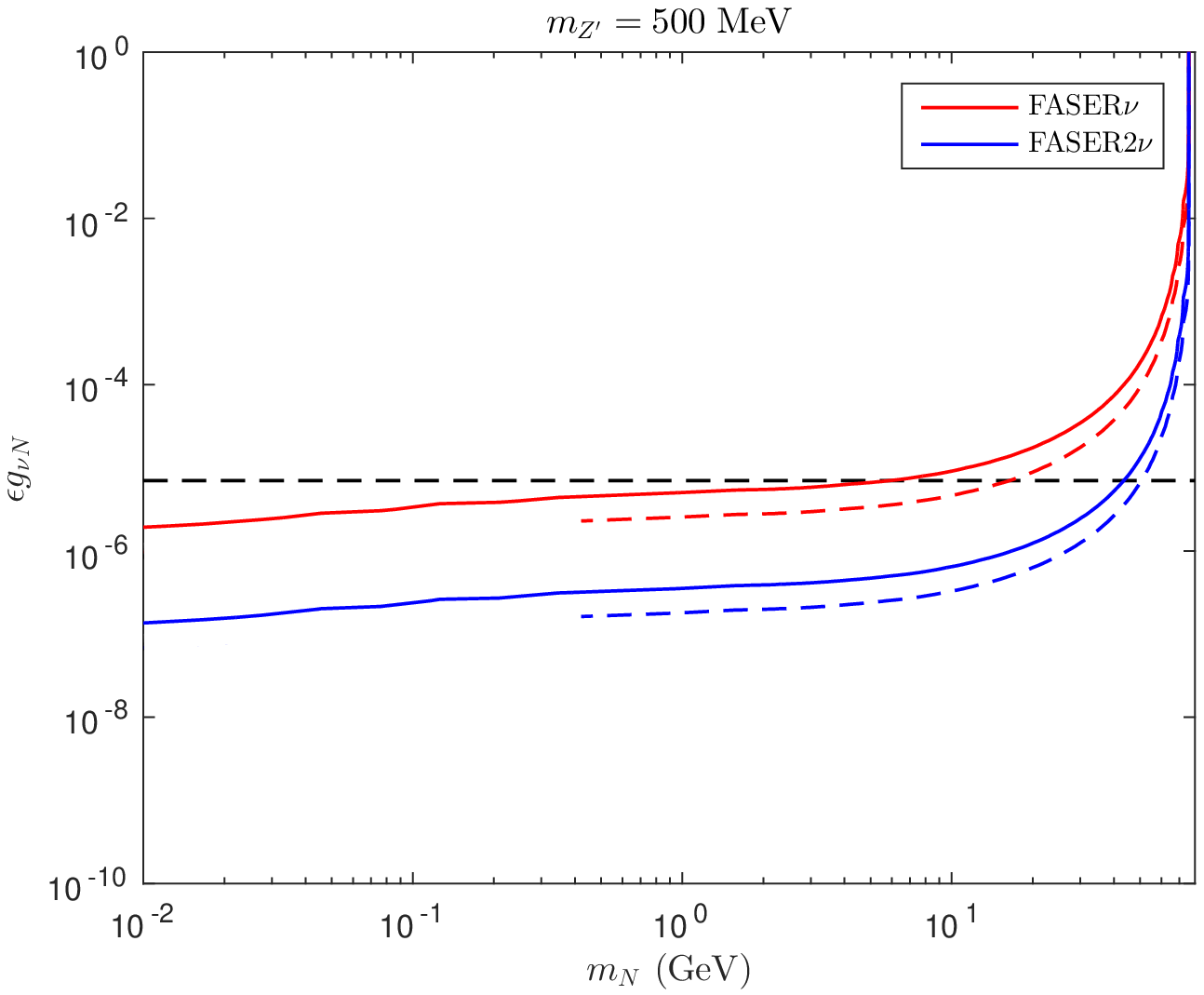}
\caption[...]{
Constraints on $\epsilon g_{\nu N}$ vs. $m_N$. In the left (right) panel, we have taken $m_{Z^\prime}$ equal to $30~\rm{MeV}$ ($500~\rm{MeV}$). The solid red curves correspond to the bound that FASER$\nu$ can obtain during run III of LHC with events originating inside the detector ({\it i.e.,} taking $N_{HESE}=1$). The dashed red curve in the right panel shows the same bound once events originating inside the rock are included ({\it i.e.,} taking $N_{HESE}+N_{through-going}=1$). The blue curves marked with FASER2$\nu$ show
the bounds that can be achieved by future upgrades of FASER$\nu$ with two hundred times more data.
The
dashed curves are clipped at $m_N=4m_\mu$ as for lighter $N$, the signal of $\mu \bar{\mu}\mu \bar{\mu}$ for through-going muons cannot be obtained.
The cyan and magenta curves correspond to the current constraints from CHARM II and MINER$\nu$A \cite{Arguelles:2018mtc}. The green area shows the solution to the MiniBOONE anomaly presented in \cite{Bertuzzo:2018itn}. The dashed black lines show $7\times 10^{-6}$ which is a combination of the bound $\epsilon<7 \times 10^{-4}$ from BABAR \cite{Lees:2014xha} and setting $g_{\nu N}=10^{-2}$.
\label{Fig:constraints}
}
\end{figure}

Fig~\ref{Fig:constraints} shows the bound that can be set on $\epsilon g_{\nu N}$ versus $m_N$ for $10~{\rm MeV}<m_N<80~$GeV setting $m_{Z'}=30$ MeV (left panel) and $m_{Z'}=500$ MeV (right panel).
The dashed (solid) red curves show the reach of FASER$\nu$ during run III of the LHC, including (without) the through-going signal. That is to draw the red solid (dashed) curve, we have set $N_{HESE}=1$ ($N_{HESE}+N_{through-going}=1$). The blue curves marked with FASER2$\nu$ show the improvement on the bounds if the data increases by a factor of 200. Such an increase is feasible during the high luminosity phase of the LHC as the integrated luminosity will increase twenty times in this phase of the LHC and the mass of the detector is under discussion to be increased by a factor of ten to thousand times \cite{Abreu:2019yak}. The results are derived under the assumption of zero background.

The dashed horizontal black lines in Fig.~\ref{Fig:constraints} show the present combined constraint on $\epsilon$ from BABAR \cite{Lees:2014xha} and on $g_{\nu N}$ from theoretical consideration (see the appendix):
$\epsilon g_{\nu N}<7\times 10^{-6}$.
As seen from the figures, for $m_N<{\rm few }$ 10 GeV, FASER$\nu$ can probe the values of the coupling well below this combined bound. As discussed before, for $m_N<$GeV, the scenario could also be tested by MINER$\nu$A and CHARM II (as shown by the cyan and magenta solid curves taken from \cite{Arguelles:2018mtc}) but the range $2~{\rm GeV}< m_N$ will be explored by FASER$\nu$ for the first time. The green region shows the solution to the MiniBooNE anomaly \cite{Arguelles:2018mtc}. As seen from the figure, the data from the run III can only probe a small  part of the parameter space of the solution.  Upgrade of FASER$\nu$ with 200 times more data can probe the entire this region.

Notice in the model that the coupling of the $Z'$ with the standard model fermions is through a kinetic mixing with the photon, $q_u'=-2q_d'=-2q_s'=(2e\epsilon /3)$ but in the gauged $B-3L_\tau$ symmetry, $q_u'=q_d'=q_s'$. However, for $m_N,m_{Z'}>4 m_\mu$ when the $\mu \bar{\mu}\mu \bar{\mu}$ signal mode is open, the bounds on $g_{\nu N}q_q'$ from FASER$\nu$ for both models will be very similar. In the latter case, since the present combined bound on $g_{\nu N}q_q'$ is  weaker and of order of $10^{-5}$, FASER$\nu$ during run III of the LHC may be able to slightly  improve the bound.
If FASER$\nu$ finds a large number of $\mu \bar{\mu} \mu \bar{\mu} $ signal exceeding few 100 according to Figs. \ref{NOEmZp} and \ref{NOEmN}, $g_{\nu N}q'_q$ should be much larger than $7 \times 10^{-6}$ so the option of $Z'$ mixed with the photon will be ruled out, providing a hint in favor of the option of $Z'$ as the gauge boson of the local $B$ symmetry.

\section{Summary and conclusions \label{summary}}
We have studied the discovery potential of FASER$\nu$ for beyond standard model interaction of neutrinos with nuclei that leads to a multilepton signature. In this model, the interaction of the neutrino flux creates a heavier fermion, $N$ whose decay produces the multi-lepton signal. Similarly to the models proposed in \cite{Ballett:2018ynz,Bertuzzo:2018itn,Abdullahi:2020nyr},
the $\nu+{\rm nucleon}\to N+X$ scattering
takes place through the exchange of a new $U(1)$ gauge boson, $Z'$ with mass smaller than $\sim 1$~GeV and kinetically mixed with the photon. The decay of the produced $N$ can then produce lepton pair via either on-shell (for $m_{Z'}<m_N$) or off-shell (for $m_{Z'}>m_N$) $Z'$.

We have shown how to build a consistent model in which the coupling between $\nu$, $N$ and $Z'$ is relatively large (of order of $10^{-2}$), respecting all the present bounds. With such a coupling, we have found that the FASER$\nu$ detector can record between few $10^4$ to 1 events as $m_{Z'}$ varies between 10 MeV to 1 GeV.
As long as $m_N\stackrel{<}{\sim}10$~GeV, the dependence of the number of events on $m_N$ is only mild because at FASER$\nu$, the center of mass energy in the scattering process is much larger than the $N$ mass. The part of the parameter space of the model with $m_N<1$~GeV has already been probed by CHARM II, MINER$\nu$A and MiniBooNE experiments but heavier $N$ can be probed by FASER$\nu$ for the first time.
We argue that the dilepton signal ({\it i.e.,} the $e^-e^+$ or $\mu^-\mu^+$ signals coming from a single vertex) produced by the neutrino flux interaction inside FASER$\nu$ can be discriminated against the background from the pair production by photons and from charmed-induced events ({\it i.e.,} $\nu_\mu (\nu_e) \to \mu (e) +c+X$ and the subsequent $c$ decay into $\mu$ (e)), respectively, by reconstructing the invariant mass of the lepton pair and by determining the event topology. After applying the cuts, the signal will be practically background free. The bound that FASER$\nu$ can set on the coupling is shown in Fig~\ref{Fig:constraints}.

The neutrino beam before reaching the detector has to pass through 100 meters of rock and concrete. For the first time, we discussed the possibility of enlarging the effective mass of the detector to search for new physics by including the events originated in the rock. All the charged particles except for muons will be absorbed in the rock so we have focused on the multimuon signal.
Unlike the case of dimuon events originating inside the detector, the dimuon events from the rock will suffer from a large background. Equipping the front side of the detector with a scintillator plate capable of recording the timing of the event can significantly reduce the background. The interface detector which is planned to be located between FASER and FASER$\nu$ \cite{Abreu:2020ddv} may also be able to achieve this goal. However, the background for the through-going dimuon events will be still large. Unlike the case of the events starting inside the FASER$\nu$ detector, we cannot employ cuts on the invariant mass of the $\mu^- \mu^+$ pair or on the event topology to reduce the background for the dimuon signals. However, the multi-muon events with more than two muons will become background free. Fig~\ref{Fig:constraints} shows how the bounds can be improved by including the through-going events in the search for the four-muon signal.

We showed how by adding a new scalar to the model a $
\mu \bar{\mu}\mu \bar{\mu}$ signal can be achieved. In this variation of the model, the coupling of
$Z'$ to the scalar dominates over the coupling to $\mu^+\mu^-$ so the $N$ decay dominantly produces a pair of these scalars rather than the $\mu\bar{\mu}$ pair. The produced $a$ and $\bar{a}$ eventually decay into $\mu \bar{\mu}$ comprising the $\mu \bar{\mu}\mu \bar{\mu}$ signal.

Throughout the paper, we have focused on the scattering of the $\nu_\mu$ flux. In principle, $\nu_e$ and $\nu_\tau$ can also have large couplings to $N$ and $Z'$. Similar arguments can be repeated for the $\nu_e$ and $\nu_\tau$ scattering, too. However, since the flux of $\nu_e$ ($\nu_\tau$) is smaller than that of $\nu_\mu$ by a factor of 15 (1000), the bounds that can be derived on the relevant coupling will be weaker by a factor of about 4 (30) compared to the bound on the $\nu_\mu$ coupling to $Z'$.

We have also discussed the option that $Z'$ is the gauge boson of a $B-3L_\tau$ local symmetry with negligible coupling to $\mu$ and the electron. In this case, the bounds from BABAR on the $Z'$ coupling
to the SM fermion is relaxed  but  there will still be a bound of from non-standard interaction of tau neutrinos which is weaker than the BABAR bound by a factor of 5 to 10. The bound can be further relaxed if instead of $B-3L_\tau$, the $B$ symmetry is gauged. Then, new chiral doublets have to be added to the model to cancel the gauge anomalies. If the observed number of $\mu \bar{\mu}\mu \bar{\mu}$ signal events by FASER$\nu$ during the run III of the LHC exceeds a hundred, these options should be taken more seriously as for the case of $Z'$ mixed with the photon, the number of $\mu \bar{\mu}\mu \bar{\mu}$ events cannot be such large. We have shown that if the statistics are enough, the parameters of the model such as the lifetime of $a$ and its mass as well as the mass of $Z'$ can be extracted from the data. In the case of the null signal at FASER$\nu$, the model-independent  bound on $g_{\nu N}q'_q$ can be improved by two orders of magnitude.

The idea of using the through-going muons to enlarge the effective mass of the detector is quite general and can be applied in various contexts beyond the four-muon signal produced by neutrino scattering \cite{Jodlowski:2019ycu}.
 The multi-muon through-going signal may originate from  the decay of a new particle produced at IP rather than by neutrino interaction in the rock \cite{FASERhollow}. Finally, studying the through-going dimuon events will increase the $\nu_\mu+{\rm nucleon}\to \mu +c+X$ data sample. Of course, eliminating the background from the pile-up of the muons from IP requires the timing of the arrival of the muons
 comprising a multi-muon signal which in turn motivates installing a scintillator plate in front of the detector.

 \begin{acknowledgments}
This project has received funding /support from the European Union’s Horizon 2020 research and innovation programme under the Marie Skłodowska -Curie grant agreement No 860881-HIDDeN.
We would like to thank the anonymous referee for the useful remarks. YF and PB have received partial financial support from Saramadan under contract No.~ISEF/M/98223 and No.~ISEF/M/99169. YF would like also to thank the ICTP staff and the INFN node of the InvisiblesPlus network in Padova. SP would like to thank ICTP for kind hospitality during the initial phases of this work. The authors would like to thank A. Ariga for the useful information and encouragement.
 \end{acknowledgments}

\section*{Appendix}
In this section, we introduce a model for the interaction of form shown in Eq. (\ref{off-coupling}).
The idea is based on the model which was introduced in Ref. \cite{Farzan:2019xor}.
In this model, there are two left-handed sterile neutrinos $N_L$ and $N'_L$ with an off-diagonal coupling to the gauge boson of the new $U(1)$ gauge symmetry, $Z'$, as follows
\be g' \bar{N}_L \gamma^\mu N'_L Z'_\mu +H.c.\ee
This form of the interaction can be easily obtained from the gauge symmetry if we simply assign opposite charges to $\psi_1=(N_L+N'_L)/\sqrt{2}$ and $\psi_2=(N_L-N'_L)/\sqrt{2}$:
\be g'(\bar{\psi}_1\gamma^\mu \psi_1-\bar{\psi}_2\gamma^\mu \psi_2)Z'_\mu=g' (\bar{N}_L\gamma^\mu N'_L+ \bar{N}'_L \gamma^\mu N_L )Z'_\mu \ . \ee

If $N'_L$ mixes with $\nu_\mu$, an interaction of form (\ref{off-coupling})
can be achieved with $g_{\nu N}=g' U_{\mu 4}$.
Notice that we do not want $N_L$ to mix with active neutrinos as there are strong bounds on such mixing for $N$ with a mass of few 100 MeV-few GeV. As we discussed in sect \ref{model} as long as $30~{\rm MeV}< m_{N'}<70$~MeV, the bounds on the mixing with $\nu_\mu$ is much more relaxed. The difference in mixing of $N_L$ and $N'_L$ requires breaking of the $U(1)$ gauge symmetry. Let us introduce a new scalar, $\phi$ which is neutral under the standard model gauge group but, under the new $U(1)$ gauge symmetry, has a charge equal to that of $\psi_2$. Moreover, let us impose a $Z_2$ symmetry under which
$$\psi_1 \leftrightarrow \psi_2, \ \ Z' \to -Z' \ \ {\rm and} \ \ \phi \to -\phi^* \ .$$ As a result, the combinations $\phi \psi_1- \phi^* \psi_2$ and $\phi \psi_1+ \phi^* \psi_2$
are respectively even and odd under the $Z_2$ symmetry. To give a Dirac mass to $N_L$ and $N'_L$, we add $N_R$ and $N'_R$ which both are singlets of the gauge groups but have opposite $Z_2$ parities: $N'_R$ ($N_R$) is $Z_2$ even (odd). We can then write the following Yukawa couplings that preserve both the gauge symmetry and the $Z_2$ symmetry:
\be Y' \bar{N}'_R(\phi \psi_1-\phi^* \psi_2)+Y \bar{N}_R(\phi \psi_1+\phi^* \psi_2)+{\rm H.c.}\ee
Without loss of generality, we can invoke the global $U(1)$ symmetry to rephase $\phi$ and make its vacuum expectation value, $\langle \phi \rangle =v_\phi/\sqrt{2}$, real.
These terms then lead to Dirac mass terms as follows
\be m_{N'} \bar{N}'_R N'_L +m_{N} \bar{N}_R N_L +{\rm H.c.,}\ee
where $m_{N'}=Y' v_\phi$ and $m_{N}=Y v_\phi$. Since $N'_R$ is $Z_2$ even, we can write the following Yukawa coupling:
\be Y_{R \alpha}\bar{N}'_R H^TcL_\alpha \ee
where $H$ is the standard model Higgs, $L_\alpha$ is the left-handed lepton doublet of flavor $\alpha$. Since we are mostly interested in $\nu_\mu$, we may identify $\alpha$ with $\mu$. The $Z_2$ symmetry forbids writing a similar Yukawa coupling for $N_R$ so only $N'$ mixes with $\nu_\alpha$. The induced mass term can be written as
\begin{eqnarray} \label{MASS} [\nu_\alpha^T \ \ (N'_L)^T \ \ (N_R')^\dagger c]c\left[ \begin{matrix} 0 & 0 & Y_{R \alpha }\langle H\rangle \cr 0 & 0 & m_{N'}\cr Y_{R \alpha }\langle H\rangle & m_{N'} & 0 \end{matrix} \right]
\left[ \begin{matrix} \nu_\alpha \cr N'_L \cr c(N'_R)^* \end{matrix} \right]
\end{eqnarray}
This mass matrix, which has a zero mass eigenstate (which is mainly composed of the active neutrino, $\nu_\alpha$), can be diagonalized by
\begin{eqnarray}
O=\left[ \begin{matrix}
1 & -\frac{Y_{R\alpha} \langle H \rangle}{m_{N'}} & 0 \cr
\frac{Y_{R\alpha} \langle H \rangle}{\sqrt{2} m_{N'}} & \frac{1}{\sqrt{2}}
& - \frac{1}{\sqrt{2}} \cr \frac{Y_{R\alpha} \langle H \rangle}{\sqrt{2} m_{N'}} & \frac{1}{\sqrt{2}} & \frac{1}{\sqrt{2}} \end{matrix}\right].
\end{eqnarray}
Thus, the mixing of $\nu_\alpha$ and $N'$ will be given by
$U_{\alpha 4}= {Y_{R\alpha} \langle H \rangle}/{m_{N'}}$. Notice that, unlike the minimal model when $N'_L$ directly couples to $\nu_\alpha$, in our model where the coupling is through $N'_R$, the mixing does not lead to a contribution of $m_{N'} U_{\alpha 4}^2$ to the $\nu_\alpha$ mass so the mixing can be relatively large ($U_{\alpha 4}\sim 0.1$).

Notice that the gauge and $Z_2$ symmetries allow a mass term of form
$$\frac{\mu}{2} (\psi_1^T c\psi_2 +\psi_2^T c \psi_1) +{\rm H.c.}=\frac{\mu}{2} (N_L^Tc N_L-(N_L')^Tc N_L') +{\rm H.c.}$$
This mass term should appear as the $(2,2)$ element of the mass matrix in
Eq. (\ref{MASS}) and would break the lepton number, inducing a Majorana mass for $\nu_\alpha$ proportional to $\mu$ like in the inverse seesaw mechanism \cite{10}.

The $N'$ particles with a mass of 50~ MeV and a mixing of $|U_{\mu 4}| \sim 10^{-2}$ can be produced in the supernova core. Since they will reach equilibrium with the matter, the bounds from supernova cooling do not rule out this model. Similarly, the $N'$ abundance in the early universe will be sufficiently reduced by scattering $N'+f \to \nu_\mu+f$ when the temperature drops its mass. As a result, this part of the parameter space is not restricted by cosmological data \cite{Dev}.

Remember that in order to have large signal sample at FASER$\nu$, we prefer the parameter range $m_{N'}\sim 50$ MeV, $m_N\sim 1$~GeV and $m_{Z'}/g'\sim 50$~MeV. To obtain
$m_N=Y v_\phi/\sqrt{2} \sim 1$~GeV, we need $v_\phi \stackrel{>}{\sim}$GeV.
$m_{N'}=Y' v_\phi/\sqrt{2} \sim 50$~MeV can then easily be obtained with $Y'/Y\sim 0.01$. However, $m_{Z'}^2$ also receives a contribution given by $g'^2 v_\phi^2/2$. With $m_{Z'}/g' \sim 500$~MeV, we can still have a handful of signal events at FASER$\nu$ (see Fig.~\ref{NOEmZp}).
To have $m_{Z'}/g'\sim 50$ MeV (and therefore to obtain a large number of events), there should be a cancellation. Of course, vacuum expectation values from additional scalars charged under new $U(1)$ will only increase $m_{Z'}^2$ as the contribution from each will be positive. However, if we invoke the Stuckelberg mechanism, we can obtain a negative contribution to $m_{Z'}^2$ which cancels out the positive contribution from $v_\phi^2$. The smaller $m_{Z'}$, the higher degree of fine tuning is required. The range $m_{Z'}/g'\sim$ 500 MeV seems to be more natural. Moreover,
lighter $Z'$ can only decay to the $e^-e^+$ pair not being able to produce multi-muon events.

If $\phi$ is lighter than $m_{Z'}/2$, $Z'$ can decay into $\phi \bar{\phi}$ faster than into the lepton pair because $g'\gg e q'_f$. $\phi$ can promptly decay into $\bar{N}_R' N_L'$ via the relatively large $Y'$ coupling. In principle, after the electroweak and the new $U(1)$ symmetry breaking through a coupling of form $\lambda_{\phi H} |\phi|^2|H|^2$, $\phi$ can mix with the Higgs and therefore obtain a coupling of form $\lambda_{\phi \mu} \phi \bar{\mu}\mu$. The coupling will be however suppressed by $\lambda_{\phi \mu} \sim \lambda_{\phi H} v_\phi m_\mu/m_H^2=10^{-5} \lambda_{\phi H} \ll Y'$ so the dominant decay mode of $\phi$ will be invisible $\phi \to \bar{N}_R' N_L'$. To avoid missing the signal, we can take $\phi$ to be heavier than $m_{Z'}/2$ so that the dominant decay mode of $Z'$ would be decay to the lepton pair. If we want the $Z'$ decay to produce $\mu \bar{\mu}\mu \bar{\mu}$ instead of just one pair of $\mu \bar{\mu}$, we may introduce another scalar, $a$, singlet under gauge symmetries, lighter than $m_{Z'}/2$ and mixed both with the Higgs and with $\phi$ through
\be A_{\phi a}a|\phi|^2+A_{H a}a|H|^2+H.c.\label{tril}\ee
The mixings of $a$ with $\phi $ and $H$ can approximately be written as $\alpha \sim A_{\phi a}v_\phi/m_\phi^2$ and $\beta \sim A_{H a}v/m_H^2$, respectively. As long as $g'\sin \alpha >q'_f$, the dominant decay mode of $Z'$ will be decay into an $a$ pair. The $a$ particle will have a Yukawa coupling of form $(\sqrt{2} m_f \sin \beta/v) a \bar{f}f$ to the SM fermions, $f$. For $2 m_\mu <m_a<2m_K$, the dominant decay of $a$ will be to $\mu^+\mu^-$ pair. As long as $\sin \beta \stackrel{>}{\sim} 10^{-3}$, the decay length of $a$ with an energy of few 100 GeV will be smaller than $\sim 10$~m.  
Let us now discuss the bounds on $\alpha$ and $\beta$ or equivalently on $A_{\phi a}$ and $A_{H a}$. In addition to the trilinear couplings shown in Eq. (\ref{tril}), the Lagrangian should include quartic couplings such as $\lambda_a |a|^4/2$. Nonzero $A_{H\phi}$ induces a nonzero $\langle a \rangle$. As long as $m_a\ll 0.1 v (\sin \beta/10^{-3})^{1/3}\lambda_a^{1/6}$, we can write $\langle a\rangle \simeq [-A_{Ha }v^2/(2\lambda_a)]^{1/3}\sim 0.1 v[(\sin\beta/10^{-3})/2\lambda_a]^{1/3}$. Notice that $\langle a \rangle$ is too small to destabilize the vacuum  $H$: $A_{H a}\langle a \rangle \ll m_{H}^2$. Moreover, for $A_{\phi H}<10 m_\phi^2/v$, $\langle a \rangle$ cannot destabilize the vacuum of $\phi$, either: $A_{\phi a}\langle a \rangle \ll m_{\phi}^2$.

Because of the $\beta$ mixing, the coupling of the Higgs (the mass eigenstate) to the SM particles will be suppressed by $\cos \beta\simeq 1-\beta^2/2$. The precision of the measured Higgs couplings is not enough to be sensitive  to $\beta$ smaller than $O(0.1)$. Through the $\lambda_a$ coupling, the standard model Higgs can decay to triple $a$ or double $a$ with rates of $\sim m_H \lambda_a^2 \sin^2\beta/(100 \pi^3)$ and  $\sim  \lambda_a^2 \sin^2\beta\langle a \rangle^2/(4 \pi m_H)$. 
 For $\sin \beta \sim 10^{-3}$, the decay of the $a$ particles produced by the Higgs decay at CMS and ATLAS will take place out of the detector so the signal will appear as invisible Higgs decay mode. Moreover, $H$ can decay into a $\phi$ pair with a rate of $\sim A_{H\phi}^2 \sin^2 \beta/(4\pi m_H)$. The produced $\phi$ will also appear as missing energy. For $\sin \beta\sim 10^{-3}$, the branching ratio of $H\to {\rm invisibles}$ will be much smaller than 1\% even for $\lambda_a\sim 1$ so the experimental bounds \cite{Zyla} can be readily satisfied. For $\sin \beta \gg 10^{-3}$, the $a$ particles can decay inside the ATLAS and CMS detectors giving rise to $\mu \bar{\mu}\mu \bar{\mu}$ and $\mu \bar{\mu}\mu \bar{\mu}\mu \bar{\mu}$  signals provided that $\lambda_a$ is large enough. For a given value of $\beta$, non-observation of such signal at CMS and ATLAS constrain $\lambda_a$.


\end{document}